# Brillouin Spectroscopy Reveals Mechanical Properties Beyond Hydration

Zhe Wang,[1] Maria Regato Herbella,[1] Fereydoon Taheri,[1] Maria de los Angeles De la Cruz Garcia,[1] Gaurav Dave,[1] and Christine Selhuber-Unkel[1†]

1 Institute for Molecular Systems Engineering and Advanced Materials (IMSEAM), INF 225, Heidelberg University, 69120, Heidelberg, Germany.

## ABSTRACT

Brillouin light scattering is a noncontact technique for probing the micromechanical properties of cells and tissues through the GHz-frequency longitudinal modulus. Interpreting Brillouin spectra in hydrated biological materials remains challenging because the Brillouin shift is influenced not only by water content but also by polymer network mechanics, relaxation processes, and viscous dissipation. Here, we investigated two model hydrogels with distinct network chemistry, together with a binary solvent mixture, to disentangle these contributions. Time- and space-resolved measurements of a hydrating hydrogel further show that Brillouin microscopy captures local mechanical changes that bulk rheometry only partially resolves. Experiments on ethanol-water mixtures further demonstrate that the Brillouin shift does not vary monotonically with hydration but instead reflects changes in the mixture's mechanical properties. In addition, by comparing longitudinal, bulk, and shear viscosities, we identify bulk viscous dissipation as an important contributor to the Brillouin response. Together, these results show that hydration modulates Brillouin spectra but does not fully determine them. Our findings provide a mechanical framework for interpreting Brillouin measurements in hydrated and biomolecular systems.

# I. INTRODUCTION

Mechanical properties are not only critical to the physiological function of living cells and tissues but also essential to regulate at the cellular level [1]. These properties arise from several factors, including the cytoskeleton, the extracellular collagen-elastin network, and cellular water content [2]. While classical mechanical methods, such as rheometer and nano indenter, have provided fundamental insights, the results often lack the spatial resolution to probe micromechanical properties at subcellular scales. Atomic force microscopy (AFM) offers nanometre-scale resolution, but it requires direct contact with the sample, which limits its use on living, adherent cells. Thus, Brillouin light scattering (BLS) has emerged as a promising optical technique to address this gap, offering non-invasive measurements of viscoelasticity through interactions with acoustic phonons [3-8]. As an all-optical confocal system, Brillouin microscope (BM) is uniquely capable of detecting internal mechanical properties in living cells [9,10]. In fact, numerous studies demonstrate that BM serves as a novel mechanical analysis method, providing unique contrast for imaging living cells, organisms, tissue sections, and the cornea [10-13]. Notably, Brillouin microscopy has also revealed that cancer spheroid core softening accompanies unjamming-driven phase transitions, independent of hydration state [14].

Despite this promise, growing attention has focused on how Brillouin scattering frequencies should be interpreted, and in particular whether the high-frequency longitudinal moduli measured by BM can substitute for low-frequency mechanical characterization methods such as AFM and rheometry in biological and medical contexts. On the one hand, in many hydrogel and tissue systems, Brillouin frequencies in hydrogels exhibit strong correlations with results from traditional mechanical testing [15]. In addition, correlations between shear and longitudinal moduli have also been observed in several biological systems [16]. On the other hand, it has also been reported that Brillouin frequency shifts can be insensitive to differences in Young's modulus among polyethylene oxide (PEO) samples of different molecular weight when the water content is matched [17]. Taken further, this observation has led to the suggestion that, in hydrated materials, the Brillouin signal is governed primarily by the water content rather than by the stiffness of the underlying polymer network [18]. Whether this view holds generally, and how hydration and network mechanics each contribute to the Brillouin response, remain open questions that are central to the interpretation of Brillouin microscopy in hydrated biological materials.

Resolving this apparent contradiction requires a deeper understanding of what BM actually measures and what that measurement means biologically. In this work, we investigate the mechanical and Brillouin responses of hydrogels with systematically varied water content to clarify the physical origin and interpretation of Brillouin measurements. We first show that the Brillouin frequency shift correlates linearly with water content in gelatin hydrogels, consistent with previous reports. However, a second, covalently crosslinked hydrogel system exhibits an opposite, hydration-induced stiffening response, demonstrating that BM tracks the network's intrinsic mechanical properties rather than water content itself. Hydration acts as a driver of mechanical change, not as the quantity directly measured. We further probed this distinction dynamically, using time- and space- resolved BLS to monitor the hydration front propagating through a hydrogel pillar and comparing it with bulk rheological measurements. It confirmed that Brillouin scattering captures local mechanical changes with a spatial and temporal resolution unavailable to conventional rheology. Finally, we address the physical significance of the Brillouin linewidth using ethanol–water mixtures as a model system [19].

## II. Theoretical background

Brillouin scattering is the inelastic scattering of light by thermally driven density fluctuations propagating at the sound velocity of the medium. The resulting Brillouin frequency shift $v_{\mathrm{B}}$ is directly related to the speed of sound $u$ in the material and the angle of the scattered light $\theta$, which is

$$v_{\mathrm{B}} = \frac{2n}{\lambda} u \, sin\left(\frac{\theta}{2}\right), \tag{1}$$

where $n$ is the refractive index, $\lambda$ is the wavelength [6]. The longitudinal storage modulus $M'$ is related to the acoustic velocity through

$$M' = \rho u^2 = \rho\left(v_{\mathrm{B}} \frac{\lambda}{2n \sin(\theta/2)}\right)^2, \tag{2}$$

where $\rho$ is the density. The corresponding longitudinal loss modulus $M''$ is related to the linewidth (full width at half maximum) $\Gamma_B$ according to [15]:

$$M'' = \rho v_B \Gamma_B \left(\frac{\lambda}{2n\sin(\theta/2)}\right)^2. \tag{3}$$

The longitudinal viscosity then can be estimated by [20,21]:

$$\eta_{long} = \frac{M''}{\omega_B} = \frac{\rho\Gamma_B}{8\pi}\left(\frac{\lambda}{n\sin(\theta/2)}\right)^2. \tag{4}$$

Equ. (4) attributes the full Brillouin linewidth to viscous relaxation. In general, the linewidth also contains a thermal-diffusion (Landau–Placzek) contribution associated with the ratio of specific heats ($C_P/C_v$), which is neglected here. This approximation is standard in the Brillouin literature for viscous liquids. For Newtonian liquids, the viscosity is independent of frequency, so the viscosity measured by BM should agree with that in low frequency. However, it is also important to note that longitudinal viscosity and shear viscosity are not equal. To understand the distinction, we consider the Navier–Stokes equation for incompressible Newtonian fluids:

$$\rho\left[\frac{\partial \boldsymbol{u}}{\partial t} + (\boldsymbol{u}\cdot\nabla)\boldsymbol{u}\right] = -\nabla P + \eta_s \Delta \boldsymbol{u}, \tag{5}$$

where $P$ is the pressure and $\eta_s$ is the shear viscosity. This equation expresses Newton's second law for a viscous fluid, describing the motion of a liquid under applied forces. The last term represents viscous stress. In this form, the liquid is assumed to be incompressible, which is appropriate for low-frequency shear rheology where,

$$\nabla\cdot\boldsymbol{u} = 0. \tag{6}$$

But this is not the case especially for high frequency. For a compressible Newtonian liquid, the Navier-Stokes equation can be written as [22]

$$\rho\left[\frac{\partial \boldsymbol{u}}{\partial t} + (\boldsymbol{u}\cdot\nabla)\boldsymbol{u}\right] = -\nabla P + \eta_s \Delta \boldsymbol{u} + \left(\eta_b + \frac{4}{3}\eta_s\right)\nabla(\nabla\cdot\boldsymbol{u}). \tag{7}$$

The last term represents a viscous force that resists spatially non-uniform compression or expansion of the fluid, arising when different regions change volume at different rates. This contribution involves two viscosity coefficients. One is the familiar shear viscosity $\eta_s$ and the other one is bulk viscosity $\eta_b$, also known as second viscosity, volume viscosity or expansion coefficient of viscosity [23-25]. In liquids and gases, molecules possess translational, rotational, and vibrational degrees of freedom. The shear viscosity $\eta_s$ is associated primarily with translational motion. In contrast, the bulk viscosity $\eta_b$ arises from the relaxation of rotational and vibrational degrees of freedom. These effects combine to define the longitudinal viscosity, which equals the sum of the bulk viscosity and four-thirds of the shear viscosity, as shown below:

$$\eta_{long} = \eta_b + \frac{4}{3}\eta_s. \tag{8}$$

Brillouin scattering probes mechanical properties at GHz frequencies, and the viscosities in the linewidth relation are therefore evaluated at these frequencies. For a Newtonian fluid, defined by a viscosity that is independent of both strain rate and frequency, the value measured at GHz frequencies is identical to that measured at low frequencies by conventional rheology. The validity of this assumption is governed by the structural relaxation time $\tau$. For simple liquids such as water and ethanol, $\tau$ (~ 1 ps) approximately two orders of magnitude shorter than the acoustic oscillation period at the Brillouin frequency (~ 130 ps at 7.5 GHz) [26]. The corresponding Deborah number De ~ 0.05 $\ll$ 1 confirms that both liquids reside in the hydrodynamic regime at GHz frequencies, where shear viscosity is effectively frequency-independent and consistent with low-frequency rheological measurements.

# 3. MATERIALS AND METHODS

### 3.1 Hydrogel preparation

Gelatin from porcine skin type A powder (G1820), 2-Hydroxyethyl methacrylate 99% (477028), Methacrylic acid 99% (155721), Tri(ethylene glycol) dimethacrylate 95% (261548) and, 2-Hydroxy-2-methylpropiophenone 97% (Darocur, 405655) were purchased from Sigma-Aldrich and used as received.

### *3.1.1 Synthesis of Gelatin hydrogels*

The Gelatin hydrogels with different gelatin proportion (180, 160, 140, 120, 100, 80, 60 and 40 mg/mL) were prepared dissolving the gelatin powder in 2 mL of distilled water at 60 ºC for 1 h. The warm solutions were poured into bottom-glass petri dish and left at 40 ºC for 2 h, covering them with parafilm to avoid the loss of solvent. Then the samples were placed at room temperature for 24 h before performing further characterization.

### *3.1.2 Synthesis of synthetic hydrogels*

The P[($HEMA_{0.85}$)-co-($MAA_{0.10}$)] hydrogels were molded on a silicon template of 6 mm diameter and 2 mm height by UV photopolymerization at 365 nm (4 $mW/cm^2$) for 7 min. Previously, 70 µL Tri (ethylene glycol) dimethacrylate, as crosslinker, and 20 µL of Darocur, as initiator, were added to the ink.

## 3.2 Brillouin spectroscopy

Brillouin scattering measurements were performed with a virtually imaged phased array (VIPA) -based high-resolution optical Brillouin spectrometer as shown in Fig. 1. The setup consists of a commercially available inverted confocal microscope (IX81, Olympus) coupled with a continuous-wave laser (532 nm, Cobolt Samba 100) and a 20× (NA,0.42) objective as previously described [27]. Data analysis was carried out following the methods described in the reference papers [6,28]. Each spectrum was fitted with a damped harmonic oscillator (DHO) model to extract the Brillouin frequency shift $\nu_B$ and linewidth $\Gamma_B$. The Brillouin linewidth was corrected for instrumental broadening by deconvolving the instrument response function (IRF), obtained from the measured linewidth of water, assuming a Gaussian instrumental profile and applying a root-mean-square relationship [21]:

$$\Gamma_B = \sqrt{{\Gamma_{experiment}}^2 - {\Gamma_{PSF}}^2}, \tag{9}$$

where $\Gamma_{experiment}$ and $\Gamma_{PSF}$ are the linewidth measured from water and broadening contributed by the spectrometer's own point spread function, respectively. For each measurement session, $\Gamma_{PSF}$ was determined from a water reference measured under identical optical conditions, using the known intrinsic linewidth of water $\Gamma_{\mathrm{B,water}} = 0.2728$ GHz [21].

Uncertainties on $v_{\mathrm{B}}$ and $\Gamma_{\mathrm{B}}$ for individual spectra were taken as the standard error returned by the DHO fit. For conditions measured with independent repeats, the uncertainty was instead taken as the standard error of the mean.

**Fig. 1: Setup of the Brillouin microscope.**

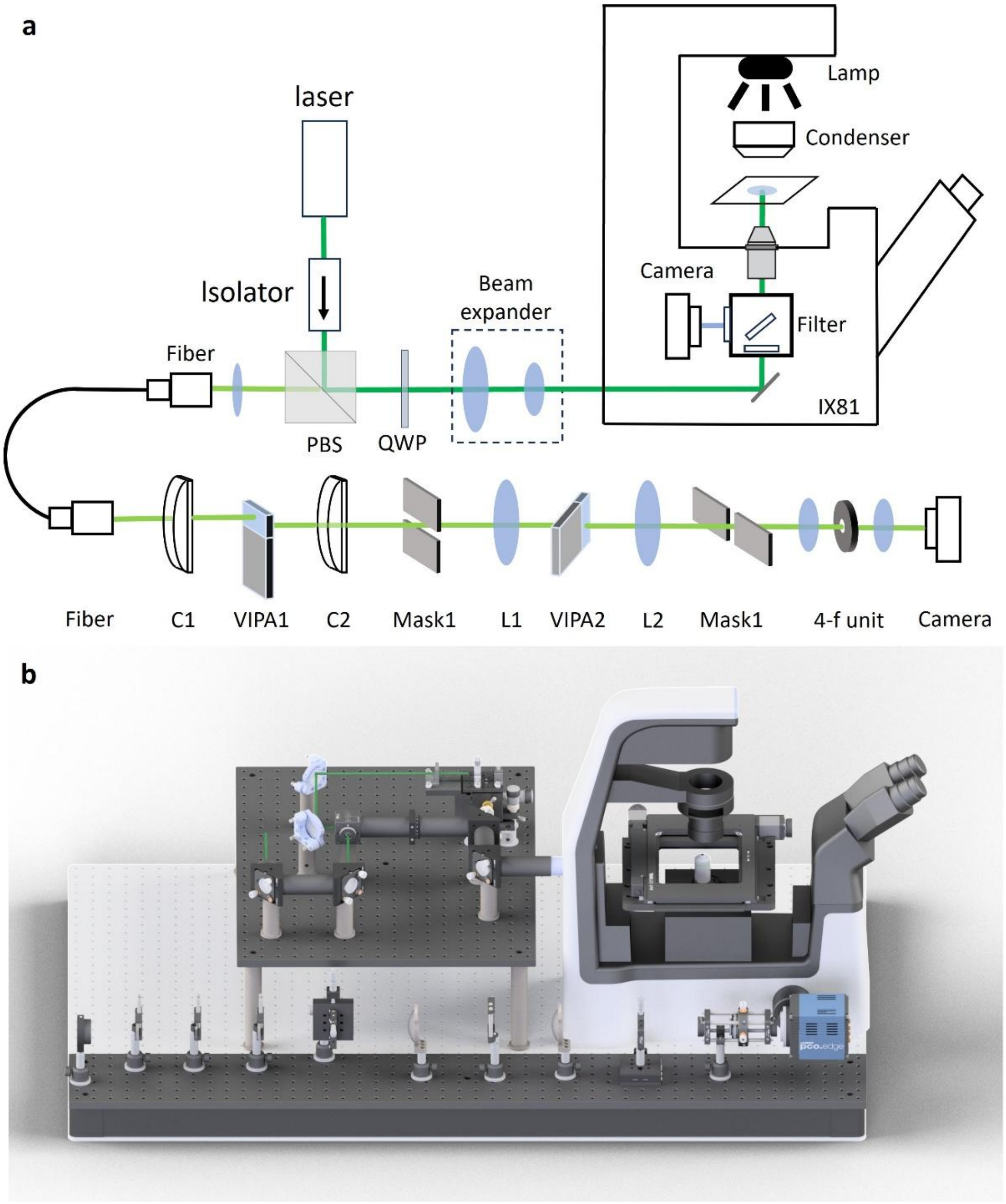

**a** Schematic of the Brillouin microscope setup. PBS, polarized beam splitter; QWP, quarter-wave plate; C1, C2, cylindrical lenes; VIPA1, VIPA2, virtually imaged phased array; L1, L2, lens while **b** is the corresponding 3D model.

### 3.3 Rheology

The shear modulus (G) was measured by a rheometer (Discovery Hybrid 20). An 8 mm diameter parallel-plate geometry was used for hydrogel samples, while liquid samples were measured using a 20 mm diameter, 0.5° cone-and-plate geometry. For hydrogels, the shear modulus was reported at 1 Hz and 1% strain at room temperature. For liquid samples, shear viscosity measurements were conducted over a frequency range from 1 to 30 kHz.

## 4. Results

To examine whether Brillouin frequency shifts primarily reflect water content or intrinsic mechanical properties, we first investigated hydrogels with systematically varied hydration. Consistent with previous reports [15], increasing water content reduced the elastic modulus and decreased the Brillouin frequency shift, with an approximately linear dependence of the shift on the water fraction over the studied range (Fig. 2a-c). The longitudinal storage and loss moduli $M'$ and $M''$ derived from the spectra (Fig. 2c), and the shear moduli measured by rheometry (Fig. 2d), follow the same overall trend, although closer inspection reveals a change of slope at high water content (~92% water), as discussed below.

The representative Brillouin peaks for gelatin hydrogels of different concentrations are shown in Fig. 2a, which indicates that the gels are homogeneous. As shown in Fig. 2b, the Brillouin frequency shift ($\upsilon_{\mathrm{B}}$) and linewidth ($\Gamma_{\mathrm{B}}$) have a good linear correlation with water content. The values of $\upsilon_{\mathrm{B}}$ and $\Gamma_{\mathrm{B}}$ were obtained by fitting the spectra with a damped harmonic oscillator (DHO) model, in line with previous observations [15,29]. Based on the reference values for refractive index $n$ and density $\rho$ [15], we further derive the corresponding longitudinal storage modulus $M'$ and loss elastic modulus $M''$ in Fig. 2c, by Eq. (2) and (3). Here $\lambda$ is excitation wavelength (532 nm) and $\theta$ is the collection angle of the scattered light, which is 180 °. A two-regime linear fit of M″ is provided at a breakpoint near 92% water content, with a steeper decrease at higher water fractions. The Brillouin (longitudinal) response is more sensitive to composition in the highly hydrated regime, where the network contribution to the GHz longitudinal modulus vanishes [30,31].

The same hydrogels were characterized by rheometry at 1 Hz in Fig. 2d, yielding the shear storage and loss modulus. At low polymer concentrations, the shear modulus remains linearly correlated with the water content. Linear fits for loss modulus are provided for two distinct regimes, below 92% and above 92% water concentration, mirroring the two-regime behaviour of M″. As the concentration increases beyond this threshold, shear modulus rises more sharply. This behavior reflects the polymer molecular arrangements in solutions. In dilute conditions, polymer chains exist as isolated coils, but at concentrations exceeding the coil overlap threshold, the chains begin to overlap and entangle, leading to a sharp increase in modulus.

The two techniques thus locate the same composition crossover (~92% water content) but respond to it in opposite ways. This difference reflects the distinct physical quantities probed, shear elasticity of the network versus the effective longitudinal response of the composite and visible in the comparison of the corresponding viscosities $\eta$. The loss modulus G″ from rheometer is related to shear viscosity through $\eta_{\mathrm{s}} = G''/\omega$, which refers to the dissipation of energy in viscous fluids. In the case of Brillouin shift, the longitudinal viscosity $\eta_{\mathrm{long}}$ can be estimated using Eq. (4) [15,18,20].

As shown in the inset of Fig. 2d, the two viscosities differ and exhibit a turning point at 92% water content, confirming that the longitudinal response measured by Brillouin scattering is not equivalent to the shear response measured by rheometry. Consistent with the findings of Wu et al., the Brillouin signal appears to be more influenced by water content than by the mechanical properties of the network in this particular experimental situation [17]. This behavior can be understood by considering that the sensitivity of Brillouin microscopy to the solid polymer fraction vanishes in highly hydrated gels. We now examine whether this water sensitivity reflects a direct contribution of water to the Brillouin signal, or the effective mechanical response of the gel.

**Fig. 2: Gelatin hydrogels Brillouin spectroscopy.**

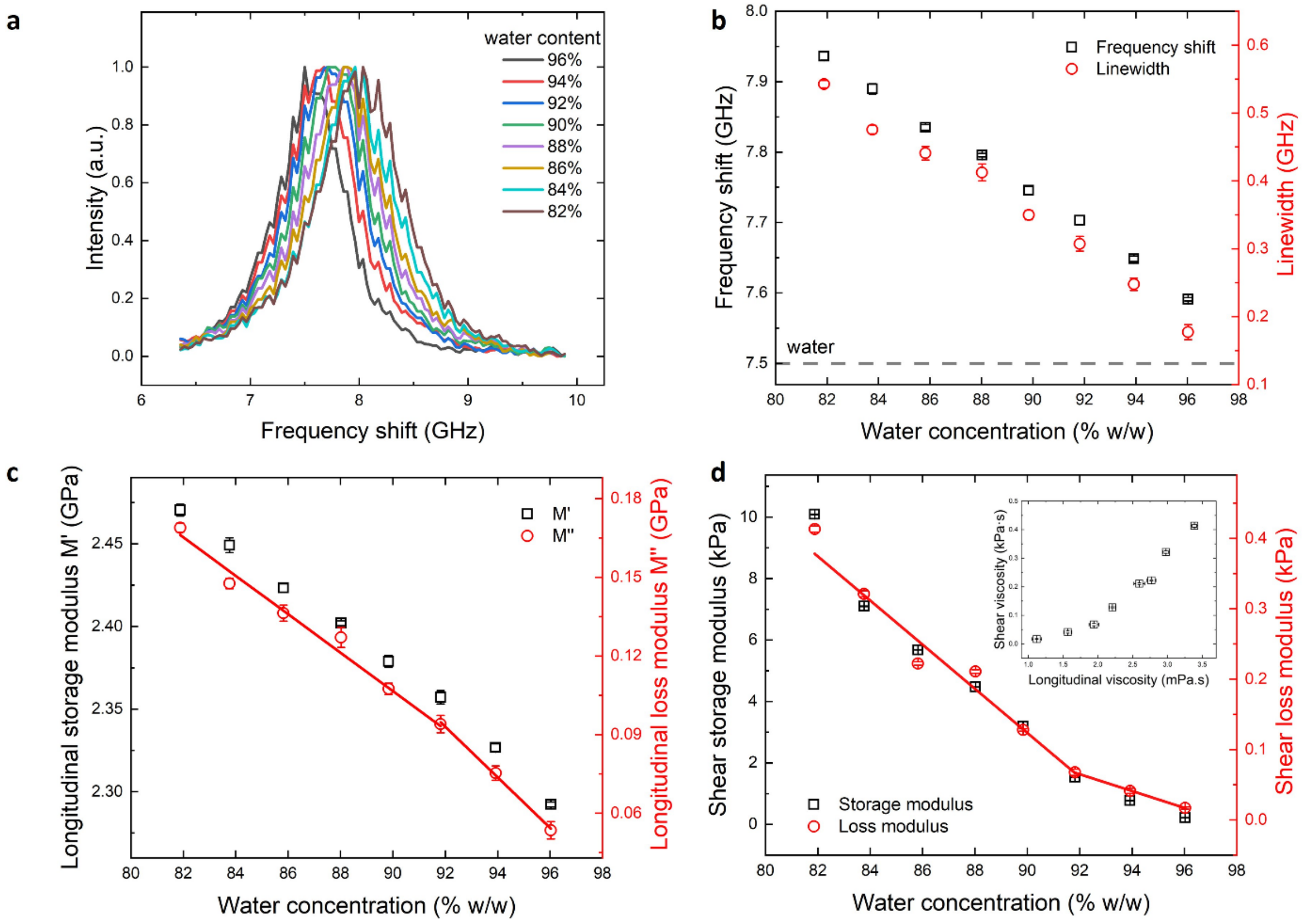

Dependence on water concentration of **a** normalized Brillouin anti-Stokes peak spectra, **b** frequency shift and linewidth (the Brillouin frequency shift of water is indicated by a dashed line), **c** storage and loss moduli extracted from Brillouin spectra (the loss modulus M″ is fitted with two linear regimes, breakpoint ≈ 92% water content), and **d** storage and loss moduli via rheometer of gelation hydrogels, the inset shows the relation between the low-frequency shear viscosity measured by rheometer and the high-frequency longitudinal viscosity inferred from the Brillouin linewidth.

As shown in Fig. 2, the Brillouin signal correlates strongly with the hydrogel's water content, with an approximately linear dependence over the studied range. This trend is intuitive while increasing water content generally leads to a softer hydrogel. Such linearity suggests that the water content can be described using a Voigt-type mixing model:

$$M_{\text{eff}} = (1 - x)M_1 + xM_{\text{water}} = (M_{\text{water}} - M_1)x + M_1, \tag{10}$$

where $M_{\text{eff}}$ is the effective elastic modulus, $M_1$ is the elastic modulus of hydrogel, $M_{\text{water}}$ is the elastic modulus of water and $x$ is the water content. As the water content increases, the Brillouin frequency shift progressively approaches that of pure water (7.5 GHz as shown in Fig. 1b), reflecting the growing contribution of the water to the overall mechanical response. Once $M_1$ and $M_{\text{water}}$ are calibrated, the water content can be estimated from this linear relationship. A qualitatively similar conclusion is reached under the Reuss limit,

$$1/M_{\text{eff}} = (1 - x)/M_1 + x/M_{\text{water}}, \tag{11}$$

which assumes uniform stress and yields a different (non-linear in $M_{\text{eff}}$) but still monotonic dependence on water content. In either bound, $M_{\text{eff}}$ is determined by x through an effective medium mixing rule, so a calibrated monotonic relationship can be used to estimate hydration from the Brillouin shift. However, this does not imply that Brillouin measurements directly quantify water content. The correlation arises from the mixture's mechanical response, not from a causal coupling to the water fraction.

In gelatin, the Brillouin shift closely follows hydration and approaches the value of pure water at high water content. This kind of behavior is often cited as evidence that Brillouin measurements primarily probe water concentration. To test the generality of this interpretation, we investigated a second hydrogel system, P[(HEMA$_{0.85}$)-co-(MAA$_{0.10}$)] with a markedly different network structure. In contrast to gelatin, chemical crosslinking in this network prevents the Brillouin shift from converging toward that of water upon hydration. Instead, the shift increases, consistent with hydration-induced network stiffening. This likely reflects swelling-induced prestress in the elastically effective chains. Osmotic and conformational contributions may also play a role, discriminating between these mechanisms is left for future work.

Fig. 3 shows the hydration behavior of cylindrical hydrogel samples (6 mm in diameter and 2 mm in height) immersed in water. Brillouin microscopy measurements were performed at the edge of the sample, with the focal point positioned at a depth of ~1 mm. Note that intense scattering occurs at the sample edge, thus, the focal point was offset slightly inward (~10 µm) rather than placed exactly at the boundary. As shown in Fig. 3a, initially, the Brillouin shift and linewidth remain nearly constant, as water has not yet penetrated to the probed region. Once water reaches the probed region, both quantities increase rapidly. We used a Kohlrausch–Williams–Watts (KWW) stretched-exponential relaxation model to describe the spatial and temporal evolution of the modulus during hydration,

$$E(x,t) = \begin{cases} E_{inf} - \left(E_{inf} - E_0\right)e^{-\left(\frac{t-\frac{x}{v}}{\tau}\right)^{\beta}} & x < vt. \\ E_0 & x > vt \end{cases} \tag{12}$$

This model was adapted to account for a hydration front propagating at constant velocity $v$. For positions already reached by the waterfront ($x < vt$), the local modulus relaxes from its unhydrated value $E_0$ toward the fully hydrated value $E_{inf}$ following a stretched exponential in the local retarded time $t - \frac{x}{v}$. The $\tau$ and $\beta$ are the characteristic timescale and stretching exponent, respectively. Ahead of the front ($x > vt$), the material remains unrelaxed. This model fitted separately to the Brillouin frequency shift and linewidth data in Fig. 3a (solid lines), with $\beta$ was fixed to 1 (i.e., a simple exponential relaxation) in both cases. For the frequency shift, the fit yielded $E_{inf} = 10.917 \pm 0.005$ GHz, $E_0 = 9.442 \pm 0.006$ GHz, and $\tau = 511 \pm 8$ s. For the linewidth, the fit gave $E_{inf} = 1.26 \pm 0.01$ GHz, $E_0 = 0.43 \pm 0.02$ GHz, and $\tau = 38 \pm 5$ s. Notably, the Brillouin frequency of pure water is 7.5 GHz. After hydration, the sample's Brillouin frequency moves even further from this value of water rather than approaching it. This behavior indicates that BLS is sensitive to hydration-induced changes in the network's mechanical properties during swelling rather than simply reflecting a mixture of water and polymer.

In Fig. 3b, spatial profiles of the Brillouin frequency shift were measured radially from the pillar edge into the interior at six timepoints during hydration (0, 15, 30, 45, 60, and 75 min), to estimate the water penetration speed. Each profile shows a sharp transition ("waterfront") separating the hydrated, higher-shift outer region from the unhydrated core. This waterfront progressively advances into the pillar over time. The inset shows the waterfront position (defined as the distance from the edge at which the frequency shift drops to the unhydrated baseline), plotted as a function of time. It indicates that a constant (linear) water penetration speed adequately describes the system over this timescale.

To estimate this speed, we performed a joint fit of all five spatial profiles (without 0 min) using Eq. (12), fixing $E_{inf}$, $E_0$, $\tau$ and $\beta$ to the values obtained from the temporal fit in Fig. 3a and treating the front velocity $v$ as a shared free parameter. Because each spatial profile was acquired at a different absolute time, an offset $t_0$ was introduced so that the effective hydration time for each profile was $t_0 + \Delta t$, where $\Delta t$ is the time elapsed since the first measurement. The fitting result (solid lines, Fig. 3b) gave a front velocity of $v = 0.0578 \pm 0.0004\ \mu\mathrm{m/s}$.

The fit describes the early timepoints (0-45 min) well, but shows increasing deviation at 60 and 75 min. This discrepancy likely arises because water penetration slows as the front advances deeper into the hydrogel, whereas the model assumes a constant velocity $v$ and a fixed characteristic timescale $\tau$ throughout. Because the corresponding linewidth data were considerably noisier and less reliable, those results are presented in the Supplementary Information rather than in the main text.

**Fig. 3: Hydration behavior of cylindrical P[(HEMA$_{0.85}$)-co-(MAA$_{0.10}$)] hydrogels samples monitored by Brillouin light scattering (BLS).**

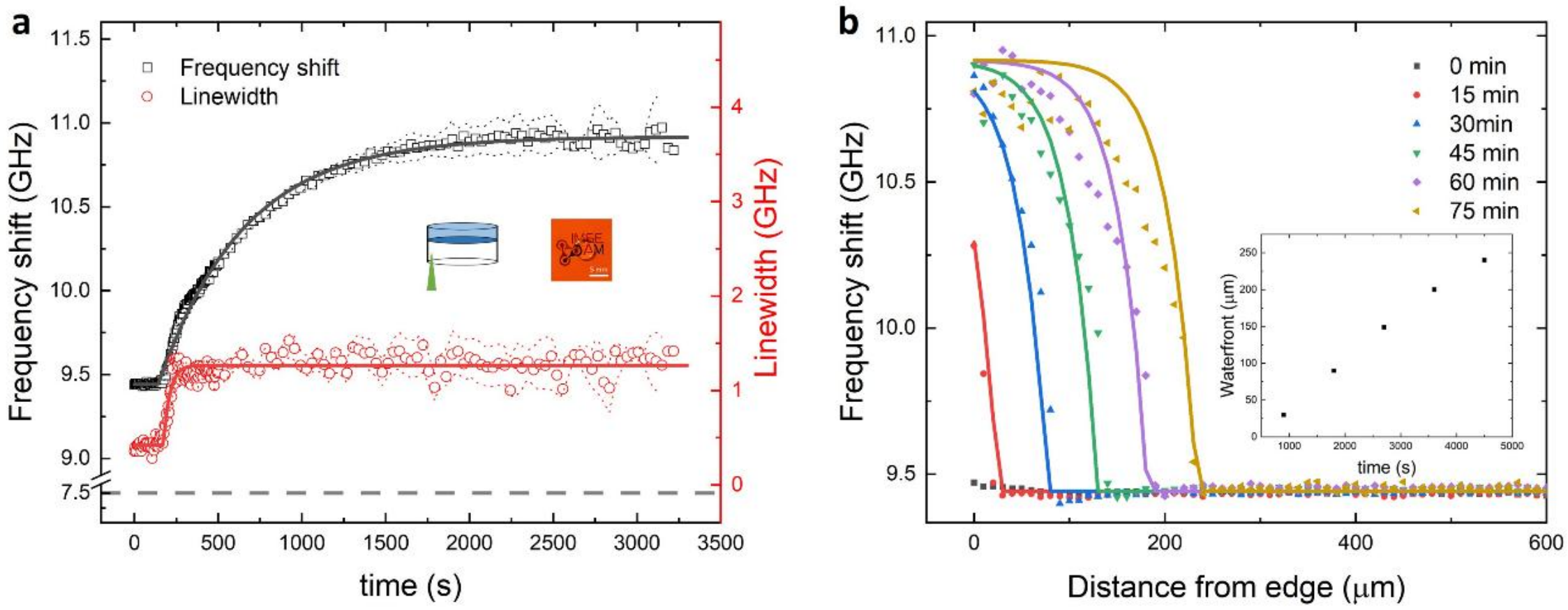

**a** Time-resolved BLS measurements acquired at the edge of the hydrogel pillar during hydration. Open squares show the Brillouin frequency shift (left axis) and open circles show the linewidth (right axis); dotted lines indicate the standard deviation. Inset: schematic of the measurement geometry, showing the optical focal point positioned at the edge of the cylindrical pillar with water penetrating from the top surface, alongside a representative image of a sample prior to immersion. **b** Spatial profiles of the Brillouin frequency shift measured radially from the pillar edge into the interior, recorded at six timepoints during hydration. Inset: waterfront position plotted as a function of time, showing the rate of water penetration into the pillar.

The corresponding shear moduli measured by rheometry are shown in Fig. 4. The bulk storage and loss moduli increase more gradually over time than the local Brillouin response shown in Fig. 3a. This difference arises because rheometry measures a volume averaged response across the entire cylindrical sample, where different layers hydrate at different times. In the case of BLS, it probes a confined focal volume at a fixed depth and therefore registers a sharp change as soon as the hydration front reaches that specific depth. This contrast highlights the ability of Brillouin microscopy to resolve spatially localized mechanical changes during swelling that are otherwise averaged out in bulk rheological measurements.

To enable a direct comparison between the two techniques, we estimated an effective Brillouin frequency shift for the whole pillar using a Voigt-type effective-medium model, in which the local shift $E(x,t)$ obtained from Fig. 3 is spatially averaged over the full sample height:

$$E_{\text{eff}}(t) = \frac{1}{L}\int_0^L E(x,t)\,dx, \tag{13}$$

where L is the height of the hydrogel pillar. Because the water penetration speed is slow and the pillar is directly attached to the Petri dish, hydration from the side and bottom surfaces was neglected. The water ingress was treated as occurring solely through the top surface. Using the front velocity and relaxation parameters obtained from the fit in Fig. 3, Eq. (13) was evaluated to generate the effective Brillouin frequency shift shown as the solid line in Fig. 4.

The effective Brillouin frequency shift correlates strongly with the storage modulus (Pearson's $r = 0.992$, $R^2 = 0.98$), more closely than with the loss modulus (Pearson's $r = 0.978$, $R^2 = 0.95$). Thus, storage modulus was chosen as the primary comparator shown in the figure. However, since rheometry and BLS operate at frequencies differing by roughly nine orders of magnitude, this correlation should not be interpreted as a general, quantitative equivalence between the two moduli. Rather, this result demonstrates that BLS is a promising tool for non-invasively monitoring hydration dynamics within hydrogels, with potential application as a mechanical biomarker for tracking drug diffusion and release in hydrogel-based delivery systems, as well as for monitoring the reversible mechanical changes in hydrogel scaffolds engineered for dynamic control of cell adhesion [32].

**Fig. 4: Correlation between rheometric shear moduli and Brillouin frequency shift during hydrogel hydration.**

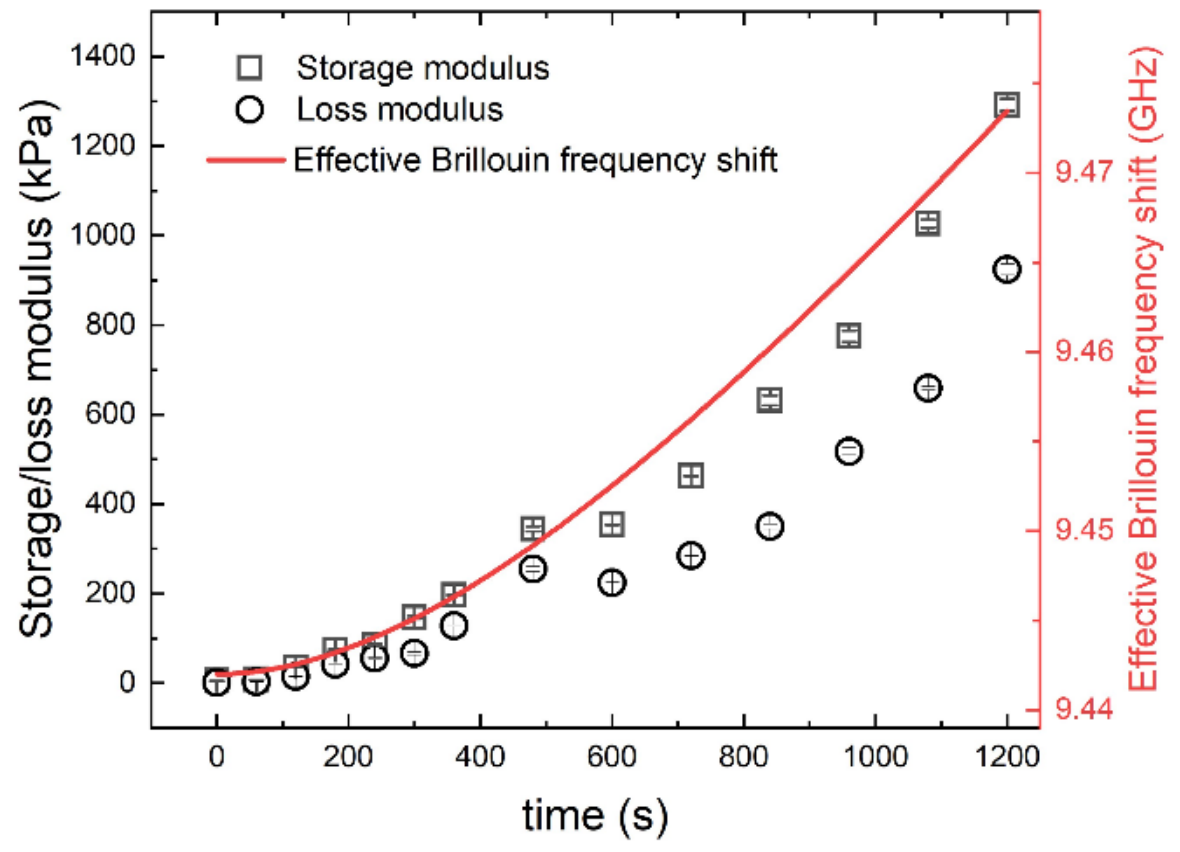

**a** Shear storage modulus (open squares) and loss modulus (open circles) measured by rheometry as a function of immersion time in water, overlaid with the effective Brillouin frequency shift (red line, right axis) obtained from the fitting parameters in Fig. 3.

To further validate that Brillouin scattering reflects the mixture's effective mechanical response rather than directly measuring water fraction, we investigated ethanol-water mixtures, which is even simpler than hydrogels. Ethanol and water are both associating liquids capable of hydrogen bonding and are miscible in all proportions. However, their longitudinal modulus and viscosity do not follow a simple linear mixing rule. These deviations arise because each component forms hydrogen-bonded networks with distinct microstructures. Individually, each liquid organizes to balance dense packing constraints with the formation of a preferred number of hydrogen bonds. Upon mixing, these competing structural tendencies interact, leading to non-ideal mechanical behavior [33].

A distinctive feature of ethanol-water mixtures is the existence of a maximum in sound velocity (proportional to frequency shift) at approximately 66 vol% water at room temperature, as confirmed in

Fig. 5a. In contrast, the Brillouin linewidth reaches its maximum at a different composition. Importantly, neither parameter varies monotonically with increasing water content, demonstrating that Brillouin scattering probes the mixture's complex mechanical response rather than simply tracking compositional changes. The linewidth reflects viscous dissipation and is directly related to the longitudinal viscosity. In compressible media, this longitudinal viscosity contains contributions from both shear and bulk viscosity, the latter corresponding to resistance against spatially non-uniform compression. Within a Newtonian framework, viscosity is frequency independent. So the shear viscosity extracted from Brillouin measurements should, in principle, be comparable to viscosities measured by conventional low-frequency methods, provided no additional relaxation processes intervene between the two frequency regimes. This condition is well satisfied for ethanol-water mixtures, whose structural relaxation times (~1 ps) are orders of magnitude shorter than the Brillouin probing timescale (~130 ps), placing the GHz-frequency acoustic response firmly within the hydrodynamic limit [26]. To quantify the relative contributions, we therefore compared the bulk and shear viscosities of ethanol-water mixtures. The shear viscosity shown in Fig. 5b was measured at 30 kHz. Additional measurements over the frequency range 1–30 kHz, presented in the Supplemental Material (SM), confirmed the expected frequency independence. The bulk viscosity was derived from the longitudinal and shear viscosity using Eq. (8). Densities of the mixed liquids were estimated using the Jouyban–Acree model, while refractive indices were obtained by interpolation from literature data [34,35]. All supporting data are provided in the SM. As shown in Fig. 5b, the bulk viscosity is comparable to or larger than the shear viscosity. At the GHz frequencies, the excess hypersonic absorption in water is known to arise from structural relaxation of the hydrogen-bond network [36]. This is particularly relevant in biomolecular systems, where bulk and hydration-shell water exhibit multiple, decoupled relaxation processes that a single scalar viscosity cannot capture [37]. Bulk viscosity therefore represents a distinct, non-negligible contribution when interpreting Brillouin measurements of hydrated biomolecular systems.

**Fig. 5: Brillouin spectroscopy and viscosity of ethanol-water mixtures.**

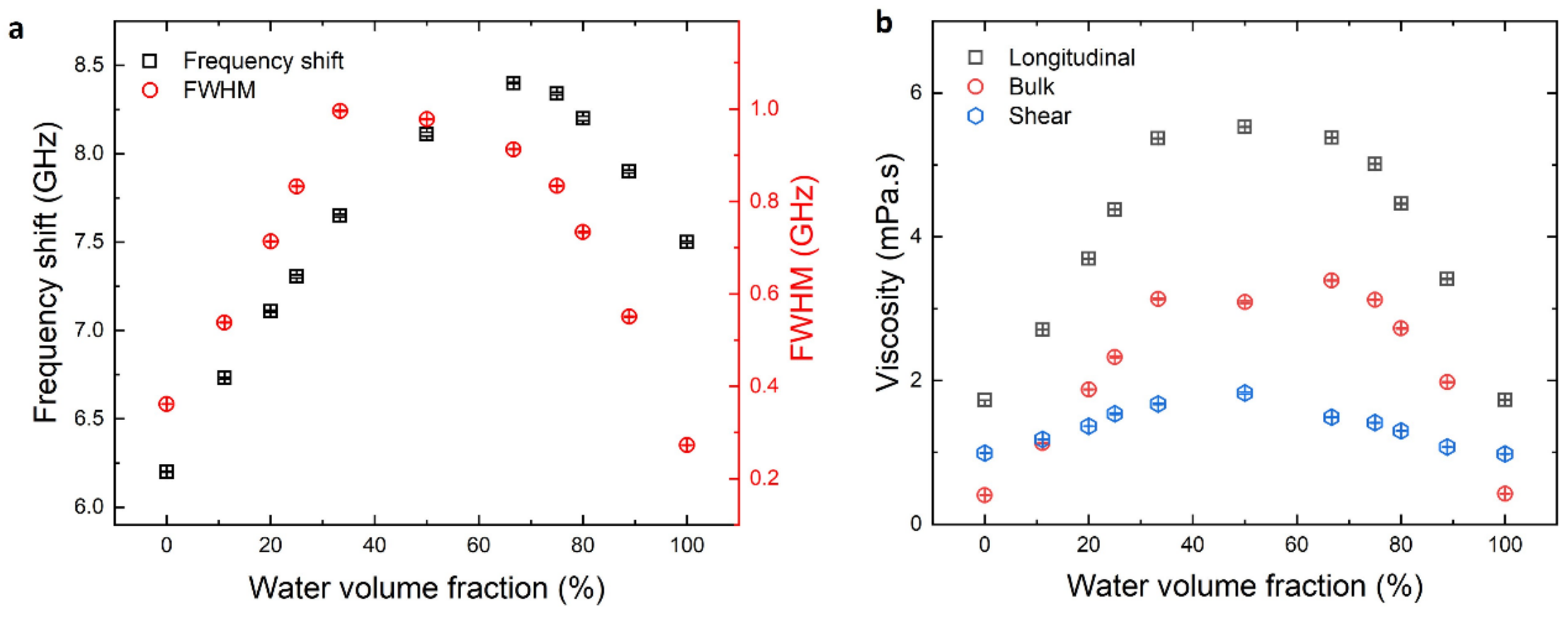

**a** Measured results of the Brillouin shift and the Brillouin linewidth in ethanol-water mixtures; **b** estimated longitudinal, bulk, and shear viscosities of the mixtures.

# 5. CONCLUSIONS

Previous studies have established important correlations between Brillouin scattering, hydration, and viscous responses in soft and biological materials, substantially advancing the interpretation of Brillouin microscopy in these systems [2,17,21]. Our findings demonstrated that Brillouin spectra are influenced by hydration, they do not simply track it. By systematically tuning gelatin formulations and solvent compositions, we demonstrated that Brillouin spectral features can either positively or negatively correlate with hydration, depending on the underlying mechanical response. Time- and space- resolved measurements of a hydrating hydrogel pillar further confirmed that Brillouin microscopy tracks local mechanical changes that are only partially captured by bulk rheometry. It provides a unique solution for monitoring dynamic processes such as swelling and, potentially, drug release in hydrogel-based delivery systems. Measurements on ethanol-water mixtures provide an independent and well-defined validation. It confirms that changes in Brillouin frequency shifts arise primarily from variations in mechanical properties rather than water content alone. These results reconcile previously divergent observations regarding the role of water content in Brillouin measurements and indicate that the GHz-frequency response reflects the underlying mechanical properties of the polymer network rather than water content alone. Characterizing the frequency dependence of the elastic and viscous response in each system remains an important direction for future work. Furthermore, by explicitly comparing longitudinal, bulk, and shear viscosities, we showed that the longitudinal response probed by Brillouin scattering incorporates significant contributions from bulk viscosity that are not captured by shear viscosity measurements alone. In particular, we find that Brillouin-derived longitudinal viscosity includes a significant bulk viscosity component, which is essential for interpreting the micromechanics of cells, tissues, and other complex biological systems [38].

## Acknowledgements

Z. W., R. M. and C. S. acknowledge the Funding from the European Research Council (ERC CoG no. 101001797, PHOTOMECH). F.T. and C.S. acknowledges support by the Deutsche Forschungsgemeinschaft (DFG; German Research Foundation) through SPP SE 1801/5-1 and the Volkswagenstiftung. A. G. and C. S. thank the European Union through funding of the Horizon project NANOBEAT (no. 101177911). G. D. and C. S. would like to acknowledge funding by the DFG under Germany's Excellence Strategy 2082/1-390761711 (3D Matter Made to Order), the Flagship Initiative "Engineering Molecular Systems." And to the Carl Zeiss Foundation through the "Carl-Zeiss Foundation-Focus@HEiKA. We thank the Core Facility "Soft (Bio)Materials Characterization" for measurement support at IMSEAM (Heidelberg University).